\RequirePackage{lineno}

\documentclass[aps,prl,reprint,lengthcheck,showpacs,superscriptaddress,nofootinbib,twocolumn]{revtex4-1}
\pdfoutput=1
\usepackage{atlasphysics}

\usepackage{subfigure}
\usepackage{mathrsfs}
\usepackage{amsmath}
\usepackage{slashed}
\usepackage{setspace}
\usepackage{footnote}
\usepackage{graphicx}
\usepackage{comment}
\usepackage[hyperindex,breaklinks]{hyperref}

\usepackage{preprintcover}

\PreprintCoverPaperTitle{Measurement of Spin Correlation in Top--Antitop Quark Events
  and Search for Top Squark Pair Production in
$pp$ Collisions at ${\sqrt{\bf s}}{ {\bf =} {\bf 8}}
$~Te\hspace{-0.05cm}V Using the ATLAS Detector}

\PreprintIdNumber{CERN-PH-EP-2014-257}
\PreprintCoverAbstract{
A measurement of spin correlation in \ttbar\ production is presented
using data collected with the ATLAS detector at the Large Hadron Collider
in proton--proton collisions at a center-of-mass energy of $8$~\tev,
corresponding to an integrated luminosity of $20.3$~fb$^{-1}$. The
correlation between the top and antitop quark spins is extracted from
dilepton \ttbar\ events by using the difference in azimuthal angle
between the two charged leptons in the laboratory frame. In the
helicity basis the measured degree of correlation corresponds to
$A_{\rm helicity} = 0.38 \pm 0.04$, in agreement with the Standard
Model prediction. A search is performed for pair production of top
squarks with masses close to the top quark mass decaying to
predominantly right-handed top quarks and a light neutralino, the
lightest supersymmetric particle. Top squarks with masses between the
top quark mass and 191~GeV are excluded at the 95\% confidence level.
}
\PreprintJournalName{PRL}

\begin{document}

\title{Measurement of Spin Correlation in Top--Antitop Quark Events
  and Search for Top Squark Pair Production in
$pp$ Collisions at ${\sqrt{\bf s}}{ {\bf =} {\bf 8}}
$~Te\hspace{-0.05cm}V Using the ATLAS Detector}

\author{The ATLAS Collaboration}

\begin{abstract}
A measurement of spin correlation in \ttbar\ production is presented
using data collected with the ATLAS detector at the Large Hadron Collider
in proton--proton collisions at a center-of-mass energy of $8$~\tev,
corresponding to an integrated luminosity of $20.3$~fb$^{-1}$. The
correlation between the top and antitop quark spins is extracted from
dilepton \ttbar\ events by using the difference in azimuthal angle
between the two charged leptons in the laboratory frame. In the
helicity basis the measured degree of correlation corresponds to
$A_{\rm helicity} = 0.38 \pm 0.04$, in agreement with the Standard
Model prediction. A search is performed for pair production of top
squarks with masses close to the top quark mass decaying to
predominantly right-handed top quarks and a light neutralino, the
lightest supersymmetric particle. Top squarks with masses between the
top quark mass and 191~GeV are excluded at the 95\% confidence level.
\end{abstract}

\maketitle

\newcommand{\alphalp}{\ensuremath{\alpha_\ell P}}
\newcommand{\alphaip}{\ensuremath{\alpha_i P}}
\newcommand{\alphalpcpc}{\ensuremath{\alpha_\ell P_{\mathrm{CPC}}}}
\newcommand{\alphalpcpv}{\ensuremath{\alpha_\ell P_{\mathrm{CPV}}}}
\newcommand{\ljets}{$\ell+$jets}
\newcommand{\ejets}{$e+$jets}
\newcommand{\zjets}{$Z/\gamma^*$+jets}
\newcommand{\mujets}{$\mu+$jets}
\newcommand{\mcatnlo}   {\mbox{\textsc{MC@NLO}}}
\newcommand{\pythia}   {\mbox{\textsc{Pythia}}}
\newcommand{\herwig}   {\mbox{\textsc{Herwig}}}
\newcommand{\jimmy}   {\mbox{\textsc{Jimmy}}}
\newcommand{\alpgen}   {\mbox{\textsc{Alpgen}}}
\newcommand{\powheg}   {\mbox{\textsc{Powheg-Box}}}
\newcommand{\sherpa}   {\mbox{\textsc{Sherpa}}}
\newcommand{\madgraph}  {\mbox{\textsc{Madgraph}}}
\newcommand{\acer}   {\mbox{\textsc{Acer}}}
\newcommand{\ppm}{$\pm$}
\newcommand\mll{\ensuremath{m_{\ell\ell}}}
\newcommand{\etmiss}{$\et^{\textrm{miss}}$}
\newcommand\htlep{\ensuremath{\HT}}
\newcommand\dbq{$\Delta\phi(\ell,b)$}
\newcommand\ddq{$\Delta\phi(\ell,d)$}

\newcommand{\vect}[1]{\hat{#1}}
\newcommand{\dphi}{$\Delta \phi$}

\def\ee{\ensuremath{e^+ e^-}}
\def\mumu{\ensuremath{\mathrm{\mu^+ \mu^-}}}
\def\emu{$e^{\pm} \mu^{\mp}$}

\newcommand\T{\rule{0pt}{+3ex}}
\newcommand\B{\rule[-2ex]{0pt}{0pt}}

\newcommand\TT{\rule{0pt}{+2.3ex}}
\newcommand\BB{\rule[-1.5ex]{0pt}{0pt}}

\newcommand\TTT{\rule{0pt}{3ex}}
\newcommand\BBB{\rule[-2ex]{0pt}{0pt}}

\newcommand{\ahel}{$A_{\rm helicity} = 0.38 \pm 0.04$}
\newcommand{\fsm}{1.20 $\pm$ 0.05~(stat) $\pm$ 0.13~(syst)}

\newcommand{\mtop}{\mbox{$m_t$}}
\newcommand{\stopi}{\mbox{$\tilde{t}_1$}}
\newcommand{\stopibar}{\mbox{$\bar{\tilde{t}}_1$}}
\newcommand{\stopistopibar}{\mbox{$\tilde{t}_1\bar{\tilde{t}}_1$}}

\newcommand{\xiz}{\mbox{$\tilde{\chi}_1^0$}}
\newcommand{\mstop}{\mbox{$m_{\tilde{t}_1}$}}
\newcommand{\xstop}{\mbox{$\sigma_{\tilde{t}_1\tilde{t}_1}$}}
\newcommand{\rstop}{\mbox{$R_{\tilde{t}_1\tilde{t}_1}$}}
\newcommand{\mxiz}{\mbox{$m_{\tilde{\chi}_1^0}$}}

\newcommand{\mstopobs}{191} % observed SUSY stop mass exclusion -1 sigma
\newcommand{\mstopexp}{178} % expected SUSY stop mass exclusion (no -1 sigma)

\DeclareGraphicsExtensions{}
\thispagestyle{empty}
\setcounter{page}{1}

%%%%%%%%%%%%%%%%%%%%%%%%%%%%%%%%%%%%%%%%%%%%%%%%%%%%%%%%%%%%%%%%%%%%%%
%%%%%%%%%%%%%%%%%%%%%%%%%%%%%%%%%%%%%%%%%%%%%%%%%%%%%%%%%%%%%%%%%%%%%%

%%%%%%%%%%%%%%%%%%%%%%%%%%%%%%%%%%%%%%%%%%%%%%%%%%%%%%%%%%%%%%%%%%%%%%
% Introduction
%%%%%%%%%%%%%%%%%%%%%%%%%%%%%%%%%%%%%%%%%%%%%%%%%%%%%%%%%%%%%%%%%%%%%%

Detailed studies of the correlation of the spin of top and antitop
quarks in \ttbar\ events produced at hadron colliders are of great
interest; they provide important precision tests of the predictions of
the Standard Model (SM) and are sensitive to many new physics
scenarios~\cite{Frederix:2007gi,Arai:2009cp,Kane:1991bg,Cheung:1996kc,Liu:2010zze,Hplus,Hplus_tauonic,Bernreuther:1997gs,Aaltonen:2009ke,Abazov:2009ae,Abazov:2009aa,Aad:2012tj,Chatrchyan:2012vca,Aad:2012rjx,Aad:2013hla,light_stop}.
The orientations of the top and antitop quark spins are transferred to
the decay products and can be measured directly via their angular
distributions~\cite{Beneke:2000hk,Kuhn,Barger_Ohnemus_Phillips,Kane:1991bg,Arens:1992fg,Mahlon:1995zn,Stelzer_Willenbrock,Brandenburg:1996df,Chang:1995ay,Bernreuther:1995cx,Dharmaratna_Goldstein,Mahlon:1997uc,Beneke:2000hk,Bernreuther:2001rq,Bernreuther:2004jv,Uwer:2004vp,Nelson:2005jp,Godbole:2006tq,Bernreuther:2008ju,parke_deltaphi,bernreutherpol,bernreuther_si_chromo}. The
strength of their correlation has been studied previously by the CDF
and D0 collaborations in proton--antiproton scattering at
1.98~\tev~\cite{cdfljetsspin,d0dilepspin,d0dilepmespin,d0ljetsdilepmespin}
and by the ATLAS and CMS collaborations in proton--proton scattering
at 7~\tev~\cite{atlasspin,spin7,cmsspin}.

In this Letter the first measurement of \ttbar\ spin correlation in
proton--proton collisions at a center-of-mass energy of 8~TeV is
presented. Because the polarization-analyzing power of the angular
distributions of charged leptons from top and antitop quark decays is
effectively 100\%~\cite{Jezabek:1994qs,Brandenburg:2002xr}, dilepton
final states of $ee$, $\mu\mu$ and $e\mu$ are analyzed. An observable
very sensitive to \ttbar\ spin correlation is the azimuthal angle
\dphi\ between the charged leptons~\cite{parke_deltaphi}, which is
also well measured by the ATLAS detector.

First, the measurement of \dphi\ is used to extract the spin
correlation strength $A_{\rm helicity} = \frac{{N_{\mathrm {like}}} -
  N_{{\mathrm {unlike}}}}{{N_{\mathrm {like}}} + N_{\mathrm
    {unlike}}}$, where ${N_{\mathrm {like}}}$ ($N_{{\mathrm
    {unlike}}}$) is the number of events where the top quark and top
antiquark spins are parallel (anti-parallel) with respect to the spin
quantization axis. This axis is chosen to be that of the helicity
basis, using the direction of flight of the top quark in the
center-of-mass frame of the \ttbar\ system. Second, to study a
specific model that predicts zero spin correlation, a search for
supersymmetric (SUSY) top squark pair production is performed.

At the Large Hadron Collider (LHC), the SUSY partners of the top
quark, the top squarks, could be produced in pairs. Models with light
top squarks are particularly attractive since they provide a solution
to the hierarchy
problem~\cite{Weinberg:1975gm,Gildener:1976ai,Weinberg:1979bn,Susskind:1978ms}.
In such models, the mass \mstop\ of the lighter top squark mass
eigenstate \stopi\ could be close to the mass of the top quark
\mtop~\cite{Carlos:1993yy,Barbieri:1987fn}. If the lightest SUSY
particle, the neutralino \xiz (or alternatively the gravitino), is
light and the top squark mass is only slightly larger than the top
quark mass, two-body decays $\stopi\rightarrow t\xiz$ in which the
momentum of \xiz\ is very small can predominate~\cite{light_stop}. The
masses of all other SUSY particles are assumed to be large. In SUSY
models where $R$-parity is conserved, such as the Minimal
Supersymmetric Standard Model
(MSSM)~\cite{Fayet:1976et,Fayet:1977yc,Farrar:1978xj,Fayet:1979sa,Dimopoulos:1981zb},
this could lead to $\ttbar\xiz\xiz$ intermediate states, appearing
like SM \ttbar\ production with additional missing transverse momentum
carried away by the escaping neutralinos, making traditional searches
exploiting kinematic differences as presented in
Refs.~\cite{Chatrchyan:2013xna,Aad:2012ywa,Aad:2012xqa,Aad:2012uu,Aad:2014kra,Aad:2014bva,Aad:2014qaa}
very difficult. \stopistopibar\ events can be distinguished from SM
\ttbar\ events through an increase of the measured \ttbar\ cross
section as analyzed in Ref.~\cite{xsec_emu}, and since top squarks
have zero spin, through measuring angular correlations sensitive to
spin correlation, as analyzed in this Letter.

%%%%%%%%%%%%%%%%%%%%%%%%%%%%%%%%%%%%%%%%%%%%%%%%%%%%%%%%%%%%%%%%%%%%%%
% The ATLAS Detector
%%%%%%%%%%%%%%%%%%%%%%%%%%%%%%%%%%%%%%%%%%%%%%%%%%%%%%%%%%%%%%%%%%%%%%

A description of the ATLAS detector can be found
elsewhere~\cite{atlaspaper}.
%
%
%
%%%%%%%%%%%%%%%%%%%%%%%%%%%%%%%%%%%%%%%%%%%%%%%%%%%%%%%%%%%%%%%%%%%%%%
% Sample composition and modeling
%%%%%%%%%%%%%%%%%%%%%%%%%%%%%%%%%%%%%%%%%%%%%%%%%%%%%%%%%%%%%%%%%%%%%%
%
This analysis uses proton--proton collision data with a center-of-mass
energy of $\sqrt{s} = 8$~TeV, corresponding to an integrated
luminosity of $20.3$~fb$^{-1}$.

Monte Carlo (MC) simulation samples are used to evaluate the
contributions, and shapes of distributions of kinematic variables, for
signal \ttbar\ events and for background processes not evaluated from
complementary data samples. All MC samples are processed with the
{\textsc{GEANT4}}~\cite{geant4} simulation of the ATLAS
detector~\cite{atlassim} and are passed through the same analysis
chain as data. The simulation includes multiple proton--proton
interactions per bunch crossing (pile-up). Events are weighted such
that the distribution of the average number of interactions per bunch
crossing matches that observed in data.

Samples of \ttbar\ events with SM spin correlation and without spin
correlation are generated using \textsc{MC@NLO}
v4.06~\cite{mcatnlo_1,Frixione:2003ei} interfaced to
\textsc{Herwig}~v6.520~\cite{herwig} for shower simulation and
hadronization. Both samples are normalized to the NNLO cross section
including next-to-next-to-leading-logarithm
corrections~\cite{xsec_tt,xsec_tt_2}. The \textsc{CT10} parton
distribution function (PDF) set~\cite{ct10} is used. For the sample
with no spin correlation, the parton shower simulation performs
isotropic decays of the top quarks whereas the full matrix element is
used for the generation of the SM spin-correlation sample. The top
quark mass is set to 172.5~\gev~\cite{ATLAS:2014wva}. The production
of a \ttbar\ pair in association with a $Z$ or $W$ boson is simulated
using \textsc{Madgraph}~5~\cite{madgraph} interfaced to
\textsc{Pythia}~v6.426~\cite{pythia} and is normalized to the
next-to-leading-order (NLO) quantum chromodynamics (QCD) cross
sections~\cite{alpgen}.

Backgrounds to \ttbar\ events with same-flavor dilepton final states
arise from the Drell--Yan $Z/\gamma^*$+jets production process with
the $Z/\gamma^*$ boson decaying into $e^{+}e^{-}$, $\mu^{+}\mu^{-}$
and $\tau^{+}\tau^{-}$, followed by leptonic decays of the $\tau$
leptons. They are generated using the \textsc{Alpgen
  v2.13}~\cite{alpgen} generator including leading-order (LO) matrix
elements with up to five additional partons. The \textsc{CTEQ6L1} PDF
set~\cite{cteq6} is used, and the cross section is normalized to the
next-to-next-to-leading-order (NNLO) QCD prediction~\cite{ZWPROD}.
Parton showering and fragmentation are modeled by \textsc{Herwig}, and
multiparton interactions are simulated by \textsc{Jimmy}~\cite{jimmy}. The
``MLM'' parton--jet matching scheme~\cite{mlm} is employed. Correction
factors are derived from data in $Z/\gamma^*$+jets-dominated control
regions and applied to the predicted yields in the signal region, to
account for the difference between the simulation prediction and data.

Single top quark background from associated $Wt$ production is modeled
with
\textsc{Powheg-Box}~r2129~\cite{powheg,Frixione:2007vw,Alioli:2010xd,powheg_wt}
interfaced with \pythia\ using the {\sc CT10} PDF set~\cite{ct10} and
normalized to the approximate NNLO QCD theoretical cross
section~\cite{stop}. Single-top $Zt$ and $WZt$ production is generated
by \madgraph~5 interfaced with \pythia.

The diboson ($WW$, $WZ$, $ZZ$) backgrounds are modeled using
\textsc{Sherpa}~v1.4.1~\cite{sherpa} and are normalized to the
theoretical calculation at NLO QCD~\cite{dibosonxs}.

The background arising from the misidentified and non-prompt leptons
(collectively referred to as ``fake leptons'') is determined from
a combination of MC simulation of $W$+jets events using \sherpa,
single-top events via $t$-channel exchange using \mcatnlo+\herwig,
\ttbar\ events with single-lepton final states using \mcatnlo+\herwig,
and data using a technique known as the matrix
method~\cite{matrixmethod1,matrixmethod2}.

Top squark pair-production samples are simulated using the
\textsc{Herwig++}~v2.6.1 \cite{herwigpp} generator with the CTEQ6L1
PDFs \cite{cteq6}. The top squarks are assumed to decay exclusively
via $\stopi\rightarrow t\xiz$. The corresponding mixing matrices  for
the top squarks and for the neutralinos are chosen such that the top
quark has a right-handed polarization in 95\% of the decays.

%%%%%%%%%%%%%%%%%%%%%%%%%%%%%%%%%%%%%%%%%%%%%%%%%%%%%%%%%%%%%%%%%%%%%%
% Event selection
%%%%%%%%%%%%%%%%%%%%%%%%%%%%%%%%%%%%%%%%%%%%%%%%%%%%%%%%%%%%%%%%%%%%%%

Candidate events are selected in the dilepton topology. The analysis
requires events selected online by inclusive single-lepton triggers
($e$ or $\mu$). Electron candidates are reconstructed from an isolated
electromagnetic calorimeter energy deposit matched to a
charged-particle track in the inner detector and must pass `medium
identification requirements'~\cite{electron_id}. Muon candidates were
reconstructed by combining tracks reconstructed in both the inner
detector and muon spectrometer~\cite{muon_id}. Jets are reconstructed
from clusters of adjacent calorimeter
cells~\cite{atlaspaper,JEScalibration} using the anti-$k_t$
algorithm~\cite{Cacciari2006,Cacciari2008,fastjet} with a radius
parameter $R = 0.4$. Jets originating from $b$-quarks were identified
(`tagged') using a multivariate discriminant employing the long
lifetime, high decay multiplicity, hard fragmentation and high mass of
$B$ hadrons~\cite{btagging1,btagging2}. The missing transverse
momentum ($\met$) is reconstructed as the magnitude of a vector sum of
all calorimeter cell energies associated with topological
clusters~\cite{METcalibration}. The following kinematic requirements
are made:

\begin{itemize}

\item Electron candidates are required to have transverse momemtum of
  $\pt > 25$~\gev\ and pseudorapidity~\footnote{The pseudorapidity
    $\eta$ is defined via the polar angle $\theta$ as $\eta= - \ln
    \tan (\theta/2)$~\cite{atlaspaper}.} of $|\eta| < 2.47$, excluding
  electrons from the transition region between the barrel and end-cap
  calorimeters defined by $1.37 < |\eta| < 1.52$. Muon candidates are
  required to have $\pt > 25$~\gev\ and $|\eta| < 2.5$. Events must
  have exactly two oppositely charged lepton candidates (\ee, \mumu,
  \emu).

\item Events must have at least two jets (after having removed the jet
  closest to the electron, if there are jets within a cone of $\Delta R
  =0.2$ around a selected electron) with $\pt > 25$~\gev\ and $|\eta|
  < 2.5$. At least one jet must be identified as a $b$-jet using a
  requirement in the multivariate discriminant corresponding to a 70\%
  $b$-tagging efficiency.

\item Events in the \ee\ and \mumu\ channels must satisfy $\met >
  30$~\gev\ to suppress backgrounds from Drell--Yan $Z/\gamma^*$+jets
  and $W$+jets events.

\item Events in the \ee\ and \mumu\ channels are required to have
  $m_{\ell\ell} > 15$~\gev\ (where $\ell$ indicates $e$ or $\mu$) to
  ensure compatibility with the simulated backgrounds and to remove
  contributions from $\Upsilon$ and $J/\psi$ production. In addition,
  $m_{\ell\ell}$ must differ by at least 10~\gev\ from the $Z$ boson
  mass ($m_Z = 91$~\gev) to further suppress the $Z/\gamma^*$+jets
  background.

\item For the \emu\ channel, no \etmiss\ or $m_{\ell\ell}$
  requirements are applied. In this case, the remaining background
  from $Z/\gamma^* (\ra \tau\tau)$+jets production is further
  suppressed by requiring that the scalar sum of the \pt\ of all
  selected jets and leptons is greater than 130~\gev.

\end{itemize}

The expected numbers of \ttbar\ signal and background events are
compared to data in Table~\ref{tab:signal_yields}. The expected yield
for top squark pair production with a top squark mass of 180~GeV and a
neutralino mass of 1~GeV is also shown.

%%%%%%%%%%%%%%%%%%%%%%%%%%%%%%%%%%%%%%%%%%%%%%%%%%%%%%%%%%%%%%%%%%%%%%
\begin{table}
\centering
\begin{tabular}{cc}
\hline
\hline
Process                                            & Yield                      \\
\hline
\ttbar                                             & $54000 \hphantom{i} ^{+\hphantom{0}3400} _{-\hphantom{0}3600} $ \\
\zjets                                             &  \hphantom{0}$2800 \pm 300$             \\
$tV$ (single top)                                  &  \hphantom{0}$2600  \pm 180$            \\
$t\bar{t}V$                                        & \hphantom{00}$80 \pm 11$                \\
$WW$, $WZ$, $ZZ$                                   &  \hphantom{0}$180 \pm 65$               \\
Fake leptons                                       & \hphantom{00}$780 \pm 780$              \\
\hline
Total non-\ttbar                                   &  \hphantom{0}$6400 \pm 860$                     \\
Expected                                       & $60000 \hphantom{i} ^{+\hphantom{0}3500} _{-\hphantom{0}3700}$    \\
Observed                                       & \hphantom{00}$60424$            \\
\hline
\vphantom{\Huge t} $\stopi\stopibar$                       & \hphantom{00}$7100 \pm 1100$               \\
($\mstop=180$~GeV, $\mxiz=1$~GeV)                          &          $ $                  \\
\hline
\hline
\end{tabular}
\caption{\label{tab:signal_yields}
Observed dilepton yield in data and the expected SUSY and
\ttbar\ signals and background contributions. Systematic uncertainties
due to theoretical cross sections and systematic uncertainties
evaluated for data-driven backgrounds are included in the
uncertainties.
}
\end{table}
%%%%%%%%%%%%%%%%%%%%%%%%%%%%%%%%%%%%%%%%%%%%%%%%%%%%%%%%%%%%%%%%%%%%%%

\begin{figure}[htb]
\centering
\includegraphics[width=85mm]{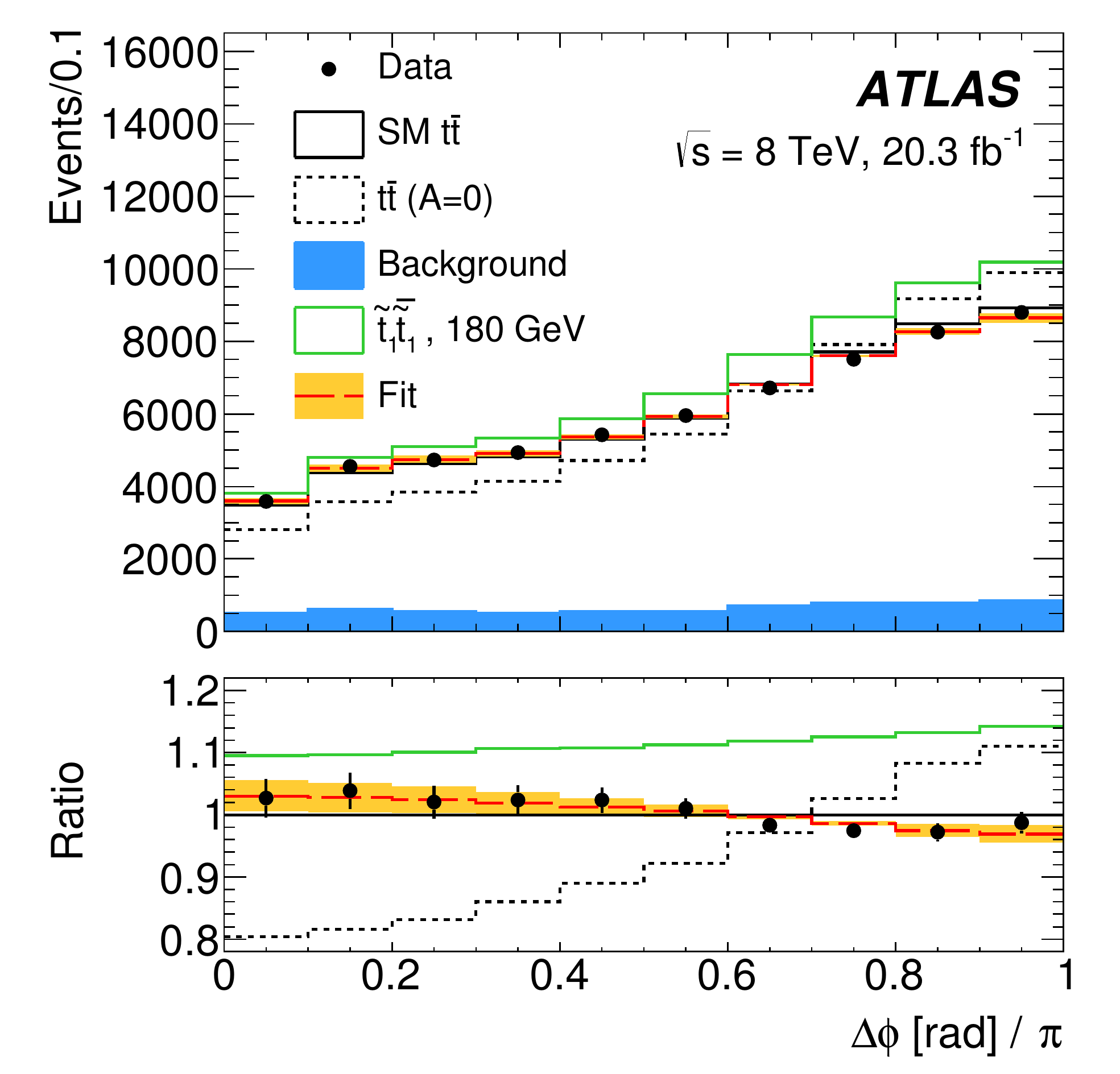}
\caption{
\label{fig:dphi_final}
Reconstructed \dphi\ distribution for the sum of the three dilepton
channels. The prediction for background (blue histogram) plus SM
\ttbar\ production (solid black histogram) and background plus
\ttbar\ prediction with no spin correlation (dashed black histogram)
is compared to the data and to the result of the fit to the data (red
dashed histogram) with the orange band representing the total
systematic uncertainty on $f_{\rm SM}$. Both the SM \ttbar\ and the no
spin correlation \ttbar\ predictions are normalized to the NNLO cross
section including next-to-next-to-leading-logarithm
corrections~\cite{xsec_tt,xsec_tt_2} (the theory uncertainty of 7\% on
this cross section is not displayed). The prediction for
\stopistopibar\ production ($\mstop =180$~GeV and $\mxiz=1$~GeV)
normalized to the NLO cross section including
next-to-leading-logarithm corrections~\cite{stopxsec} plus SM
\ttbar\ production plus background is also shown (solid green
histogram). The lower plot shows those distributions (except for
background only) divided by the SM \ttbar\ plus background
prediction.
}
\end{figure}

%%%%%%%%%%%%%%%%%%%%%%%%%%%%%%%%%%%%%%%%%%%%%%%%%%%%%%%%%%%%%%%%%%%%%%
% Measurement procedure
%%%%%%%%%%%%%%%%%%%%%%%%%%%%%%%%%%%%%%%%%%%%%%%%%%%%%%%%%%%%%%%%%%%%%%
%
Figure~\ref{fig:dphi_final} shows the reconstructed \dphi\ distribution
for the sum of the three dilepton channels.
%
%%%%%%%%%%%%%%%%%%%%%%%%%%%%%%%%%%%%%%%%%%%%%%%%%%%%%%%%%%%%%%%%%%%%%%
%%%%%%%%%%%%%%%%%%%%%%%%%%%%%%%%%%%%%%%%%%%%%%%%%%%%%%%%%%%%%%%%%%%%%%
%
A binned log-likelihood fit is used to extract the spin correlation
from the \dphi\ distribution in data. This is done by defining a
coefficient $f_{\rm SM}$ that measures the degree of spin correlation
relative to the SM prediction.
The fit includes a linear superposition of the \dphi\ distribution
from SM \ttbar\ MC simulation with coefficient $f_{\rm SM}$, and from
the \ttbar\ simulation without spin correlation with coefficient
$(1-f_{\rm SM})$. The \ee, \mumu\ and \emu\ channels are fitted
simultaneously with a common value of $f_{\rm SM}$, leaving the
\ttbar\ normalization free with a fixed background normalization.
The \ttbar\ normalization obtained by the fit agrees with the
theoretical prediction of the production cross section~\cite{xsec_tt}
within the uncertainties. Negative values of $f_{\rm SM}$ correspond
to an anti-correlation of the top and antitop quark spins.  A value of
$f_{\rm SM}=0$ implies that the spins are uncorrelated and values of
$f_{\rm SM}>1$ indicate a degree of \ttbar\ spin correlation larger
than predicted by the SM.

%%%%%%%%%%%%%%%%%%%%%%%%%%%%%%%%%%%%%%%%%%%%%%%%%%%%%%%%%%%%%%%%%%%%%%
% Systematic uncertainties
%%%%%%%%%%%%%%%%%%%%%%%%%%%%%%%%%%%%%%%%%%%%%%%%%%%%%%%%%%%%%%%%%%%%%%

Systematic uncertainties are evaluated by applying the fit procedure
to pseudo-experiments created from simulated samples modified to
reflect the systematic variations. The fit of $f_{\rm SM}$ is repeated
to determine the effect of each systematic uncertainty using the
nominal templates. The difference between the means of Gaussian fits
to the results from many pseudo-experiments using nominal and modified
pseudo-data is taken as the systematic uncertainty on $f_{\rm
  SM}$~\cite{nuisance}.

The various systematic uncertainties are estimated in the same way as
in Ref.~\cite{spin7} with the following exceptions: since this
analysis employs $b$-tagging, the associated uncertainty is estimated
by varying the relative normalizations of simulated $b$-jet, $c$-jet
and light-jet samples. The uncertainty due the choice of generator is
determined by comparing the default to an alternative \ttbar\ sample
generated with the \powheg\ generator interfaced with \pythia. The
uncertainty due to the parton shower and hadronization model is
determined by comparing two \ttbar\ samples generated by \alpgen, one
interfaced with \pythia\ and the other one interfaced with
\herwig. The uncertainty on the amount of initial- and final-state
radiation (ISR/FSR) in the simulated \ttbar\ sample is assessed by
comparing \alpgen\ events, showered with \pythia, with varied amounts
of ISR and FSR. As in Ref.~\cite{spin7}, the size of the variation is
compatible with the recent measurements of additional jet activity in
$t\bar{t}$ events~\cite{vetopaper}. The $Wt$ normalization is varied
within the theoretical uncertainties of the cross-section
calculation~\cite{stop}, and the sensitivity to the interference
between $Wt$ production and \ttbar\ production at NLO is studied by
comparing the predictions of \powheg\ with the diagram-removal
(baseline) and diagram-subtraction schemes~\cite{powheg_wt,LHC_wt}. As
in Ref.~\cite{spin7}, the uncertainty due to the top quark mass is
evaluated but not included in the systematic uncertainties, since it
would have no significant impact on the results.

The sizes of the systematic uncertainties in terms of $\Delta f_{\rm
  SM}$ are listed in Table~\ref{tab:systematics_table}.  The total
systematic uncertainty is calculated by combining all systematic
uncertainties in quadrature.

%%%%%%%%%%%%%%%%%%%%%%%%%%%%%%%%%%%%%%%%%%%%%%%%%%%%%%%%%%%%%%%%%%%%%%
\begin{table}
\centering
\begin{tabular}{cc}
\hline
\hline
Source of uncertainty           & $\Delta f_{\mathrm{SM}}$ \\
\hline
\multicolumn{2}{l}{Detector modeling} \\
\hline
Lepton reconstruction & $\pm 0.01$                \\
Jet energy scale           & $\pm 0.02$                \\
Jet reconstruction    & $\pm 0.01$                \\
\met\                       & $<0.01$                \\
Fake leptons            & $<0.01$                \\
$b$-tagging           & $<0.01$                \\
\hline
\multicolumn{2}{l}{Signal and background modeling} \\
\hline
Renormalization/factorization scale      & $\pm0.05$                \\
MC generator   & $\pm0.03$                \\
Parton shower and fragmentation     & $\pm0.06$                \\
ISR/FSR               & $\pm0.06$                \\
Underlying event       & $\pm0.04$                \\
Color reconnection   & $\pm0.01$                \\
PDF uncertainty       & $\pm0.05$                \\
Background            & $\pm0.01$                \\
MC statistics           & $\pm0.04$                   \\
\hline
Total systematic uncertainty     & $\pm0.13$                \\ %TODO
Data statistics       & $\pm0.05$                \\
\hline
\hline
\end{tabular}
\caption{\label{tab:systematics_table}
Summary of systematic uncertainties on $f_{\mathrm{SM}}$ in the
combined dilepton final state.
}
\end{table}
%%%%%%%%%%%%%%%%%%%%%%%%%%%%%%%%%%%%%%%%%%%%%%%%%%%%%%%%%%%%%%%%%%%%%%

%%%%%%%%%%%%%%%%%%%%%%%%%%%%%%%%%%%%%%%%%%%%%%%%%%%%%%%%%%%%%%%%%%%%%%
% Spin correlation strength
%%%%%%%%%%%%%%%%%%%%%%%%%%%%%%%%%%%%%%%%%%%%%%%%%%%%%%%%%%%%%%%%%%%%%%
%
The measured value of $f_{\rm SM}$ for the combined fit is \fsm. This
agrees with previous results from ATLAS using data at a center-of-mass
energy of 7~TeV~\cite{atlasspin,spin7}, and compares to the best
previous measurement using \dphi\ of $f_{\rm SM}=1.19 \pm 0.09~({\rm
  stat}) \pm 0.18~({\rm syst})$~\cite{spin7}. It also agrees with the
SM prediction to within two standard deviations.

This agrees with previous results from ATLAS using data at a
center-of-mass energy of 7~TeV~\cite{atlasspin,spin7} and agrees with
the SM prediction to within two standard deviations. An indirect
extraction of $A_{\rm helicity}$ can be achieved by assuming that the
\ttbar\ sample is composed of top quark pairs as predicted by the SM,
but with varying spin correlation. In that case, a change in the
fraction $f_{\rm SM}$ leads to a linear change of $A_{\rm helicity}$
(see also Ref.~\cite{spin7}), and a value of the spin correlation
strength in the helicity basis $A_{\rm helicity}$ at a center-of-mass
energy of 8~TeV is obtained by applying the measured value of $f_{\rm
  SM}$ as a multiplicative factor to the SM prediction of $A^{\rm
  SM}_{\rm helicity} = 0.318 \pm
0.005$~\cite{bernreuther_si_chromo}. This yields a measured value of
\ahel.

%%%%%%%%%%%%%%%%%%%%%%%%%%%%%%%%%%%%%%%%%%%%%%%%%%%%%%%%%%%%%%%%%%%%%%w
% Spin correlation in top squark pair production
%%%%%%%%%%%%%%%%%%%%%%%%%%%%%%%%%%%%%%%%%%%%%%%%%%%%%%%%%%%%%%%%%%%%%%

The measurement of the variable $\Delta \phi$ is also used to search
for top squark pair production with $\stopi\rightarrow t\xiz$
decays. The present analysis is sensitive both to changes in the yield
and to changes in the shape of the $\Delta \phi$ distribution caused
by a potential admixture of \stopistopibar\ with the SM \ttbar\
sample. An example is shown in Fig.~\ref{fig:dphi_final}, where the
effect of \stopistopibar\ production in addition to SM \ttbar\
production and backgrounds is compared to data. No evidence for
\stopistopibar\ production was found.

Limits are set on the top squark pair-production cross section by
fitting each bin of the $\Delta \phi$ distribution to the difference
between the data and the SM prediction, varying the top squark signal
strength $\mu$. In contrast to the measurement of $f_{\rm SM}$ where
the \ttbar\ cross section is varied in the fit, here the \ttbar\ cross
section is fixed to its SM value~\cite{xsec_tt}. In addition, a
systematic uncertainty of 7\% is introduced, composed of factorization
and renormalization scale variation, top quark mass uncertainty, PDF
uncertainty and uncertainty in the measurement of the beam energy.
All other sources of systematic uncertainty are identical to ones in
the measurement of $f_{\rm SM}$. All shape-dependent modeling
uncertainties on the SUSY signal are found to be negligible. The
limits are determined using a profile likelihood ratio in the
asymptotic limit~\cite{asimov}, using nuisance parameters to account
for the theoretical and experimental uncertainties.

%%%%%%%%%%%%%%%%%%%%%%%%%%%%%%%%%%%%%%%%%%%%%%%%%%%%%%%%%%%%%%%%%%%%%%w
\begin{figure}[hb]
\centering
\includegraphics[width=85mm]{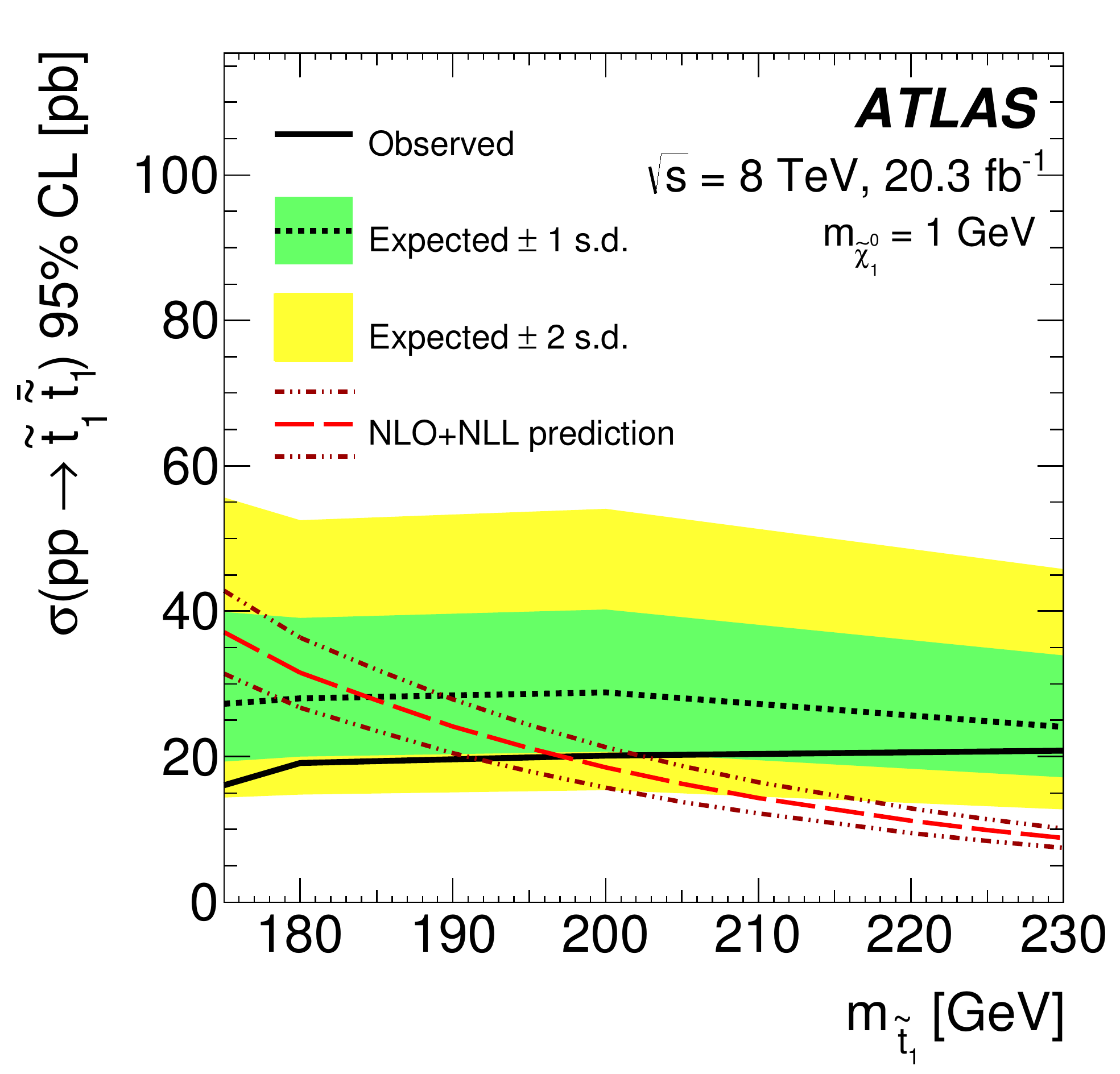}
\caption{
\label{f:susy}
Expected and observed limits at 95\% CL on the top squark
pair-production cross section as a function of \mstop, for
pair-produced top squarks \stopi\ decaying with 100\% branching ratio
via $\stopi\rightarrow t\xiz$ to predominantly right-handed top
quarks, assuming $\mxiz=1$\,\GeV. The black dotted line shows the
expected limit with $\pm 1$ (green) and $\pm 2$ (green+yellow)
standard deviation contours, taking into account all
uncertainties. The red dashed line shows the theoretical cross section
with uncertainties. The solid black line gives the observed limit.
}
\end{figure}
%%%%%%%%%%%%%%%%%%%%%%%%%%%%%%%%%%%%%%%%%%%%%%%%%%%%%%%%%%%%%%%%%%%%%%w

The observed and expected limits on the top squark pair-production
cross section at the 95\% confidence level (CL) are extracted using
the CLs prescription \cite{cls} and are shown in Fig.~\ref{f:susy}.
Adopting the convention of reducing the estimated SUSY production
cross section by one standard deviation of its theoretical uncertainty
(15\%, coming from PDFs and QCD scale uncertainties \cite{susyunc}),
top squark masses between the top quark mass and \mstopobs\,\GeV\ are
excluded, assuming a 100\% branching ratio for $\stopi\rightarrow
t\xiz$ and $\mxiz=1$\,\GeV. The expected limit is \mstopexp\,\GeV. In
the presented range of \mstop, within the allowed phase space, varying
the neutralino mass does not affect the cross-section limits by more
than a few percent. If the top quarks are produced with full
left-handed polarization, the expected limits change by less than 10\%
compared to the predominantly right-handed case.

If the \ttbar\ cross-section normalization were arbitrary and not
fixed to its theory prediction, the expected cross-section limit would
increase by approximately 30\%. If, on the other hand, the shape
information of $\Delta \phi$ were not used in the fit, the expected
cross-section limit would increase by 30-40\%.

The constraints on the top squark mass presented here improve previous
limits in a region not explored before, to top squark masses larger
than limits from Ref.~\cite{xsec_emu} and to top squark masses lower
than limits from analyses exploring kinematic distributions as
presented in Ref.~\cite{Aad:2014kra}.

%%%%%%%%%%%%%%%%%%%%%%%%%%%%%%%%%%%%%%%%%%%%%%%%%%%%%%%%%%%%%%%%%%%%%%
% Conclusion
%%%%%%%%%%%%%%%%%%%%%%%%%%%%%%%%%%%%%%%%%%%%%%%%%%%%%%%%%%%%%%%%%%%%%%

In conclusion, the first measurement of \ttbar\ spin correlation in
proton--proton scattering at a center-of-mass energy of 8~TeV at the
LHC has been presented using $20.3$~fb$^{-1}$ of ATLAS data in the
dilepton decay topology. A template fit is performed to the
\dphi\ distribution and the measured value of $f_{\rm SM}=$\fsm\ is
consistent with the SM prediction. This represents the most precise
measurement to date. The results have been used to search for
pair-produced supersymmetric top squarks decaying to top quarks and
light neutralinos. Assuming 100\% branching ratio for the decay
$\stopi\rightarrow t\xiz$, and the production of predominantly
right-handed top quarks, top squark masses between the top quark mass
and \mstopobs\,\GeV\ are excluded at 95\% CL, which is an improvement
over previous constraints.

%%%%%%%%%%%%%%%%%%%%%%%%%%%%%%%%%%%%%%%%%%%%%%%%%%%%%%%%%%%%%%%%%%%%%%
% Acknowledgements for papers with collision data
% Version 01-Mar-2014
%%%%%%%%%%%%%%%%%%%%%%%%%%%%%%%%%%%%%%%%%%%%%%%%%%%%%%%%%%%%%%%%%%%%%%

We thank CERN for the very successful operation of the LHC, as well as the
support staff from our institutions without whom ATLAS could not be
operated efficiently.

We acknowledge the support of ANPCyT, Argentina; YerPhI, Armenia; ARC,
Australia; BMWFW and FWF, Austria; ANAS, Azerbaijan; SSTC, Belarus; CNPq and FAPESP,
Brazil; NSERC, NRC and CFI, Canada; CERN; CONICYT, Chile; CAS, MOST and NSFC,
China; COLCIENCIAS, Colombia; MSMT CR, MPO CR and VSC CR, Czech Republic;
DNRF, DNSRC and Lundbeck Foundation, Denmark; EPLANET, ERC and NSRF, European Union;
IN2P3-CNRS, CEA-DSM/IRFU, France; GNSF, Georgia; BMBF, DFG, HGF, MPG and AvH
Foundation, Germany; GSRT and NSRF, Greece; RGC, Hong Kong SAR, China; ISF, MINERVA, GIF, I-CORE and Benoziyo Center, Israel; INFN, Italy; MEXT and JSPS, Japan; CNRST, Morocco; FOM and NWO, Netherlands; BRF and RCN, Norway; MNiSW and NCN, Poland; GRICES and FCT, Portugal; MNE/IFA, Romania; MES of Russia and ROSATOM, Russian Federation; JINR; MSTD,
Serbia; MSSR, Slovakia; ARRS and MIZ\v{S}, Slovenia; DST/NRF, South Africa;
MINECO, Spain; SRC and Wallenberg Foundation, Sweden; SER, SNSF and Cantons of
Bern and Geneva, Switzerland; NSC, Taiwan; TAEK, Turkey; STFC, the Royal
Society and Leverhulme Trust, United Kingdom; DOE and NSF, United States of
America.

The crucial computing support from all WLCG partners is acknowledged
gratefully, in particular from CERN and the ATLAS Tier-1 facilities at
TRIUMF (Canada), NDGF (Denmark, Norway, Sweden), CC-IN2P3 (France),
KIT/GridKA (Germany), INFN-CNAF (Italy), NL-T1 (Netherlands), PIC (Spain),
ASGC (Taiwan), RAL (UK) and BNL (USA) and in the Tier-2 facilities
worldwide.

%%%%%%%%%%%%%%%%%%%%%%%%%%%%%%%%%%%%%%%%%%%%%%%%%%%%%%%%%%%%%%%%%%%%%%
%%%%%%%%%%%%%%%%%%%%%%%%%%%%%%%%%%%%%%%%%%%%%%%%%%%%%%%%%%%%%%%%%%%%%%

%%%%%%%%%%%%%%%%%%%%%%%%%%%%%%%%%%%%%%%%%%%%%%%%%%%%%%%%%%%%%%%%%%%%%%
%%%%%%%%%%%%%%%%%%%%%%%%%%%%%%%%%%%%%%%%%%%%%%%%%%%%%%%%%%%%%%%%%%%%%%

\onecolumngrid
\clearpage
% ATLAS Collaboration author list
% Data extracted on 26-Jan-2015 for paper reference TOPQ-2014-07
%\documentclass[11pt]{article}
%\usepackage{a4wide}\begin{document}
\begin{flushleft}
{\Large The ATLAS Collaboration}

\bigskip

G.~Aad$^{\rm 85}$,
B.~Abbott$^{\rm 113}$,
J.~Abdallah$^{\rm 152}$,
S.~Abdel~Khalek$^{\rm 117}$,
O.~Abdinov$^{\rm 11}$,
R.~Aben$^{\rm 107}$,
B.~Abi$^{\rm 114}$,
M.~Abolins$^{\rm 90}$,
O.S.~AbouZeid$^{\rm 159}$,
H.~Abramowicz$^{\rm 154}$
% ...
\bigskip
\\
$^{1}$ Department of Physics, University of Adelaide, Adelaide, Australia\\
% ...
$^{11}$ Institute of Physics, Azerbaijan Academy of Sciences, Baku, Azerbaijan\\
% ...
$^{85}$ CPPM, Aix-Marseille Universit{\'e} and CNRS/IN2P3, Marseille, France\\
% ...
$^{90}$ Department of Physics and Astronomy, Michigan State University, East Lansing MI, United States of America\\
% ...
$^{107}$ Nikhef National Institute for Subatomic Physics and University of Amsterdam, Amsterdam, Netherlands\\
% ...
$^{113}$ Homer L. Dodge Department of Physics and Astronomy, University of Oklahoma, Norman OK, United States of America\\
$^{114}$ Department of Physics, Oklahoma State University, Stillwater OK, United States of America\\
% ...
$^{117}$ LAL, Universit{\'e} Paris-Sud and CNRS/IN2P3, Orsay, France\\
% ...
$^{152}$ Institute of Physics, Academia Sinica, Taipei, Taiwan\\
% ...
$^{154}$ Raymond and Beverly Sackler School of Physics and Astronomy, Tel Aviv University, Tel Aviv, Israel\\
% ...
$^{159}$ Department of Physics, University of Toronto, Toronto ON, Canada\\
% ...
\end{flushleft}

%\end{document}
% Created with ./xml2latex.py

\end{document}